An Outcome Model Approach to Translating a Randomized Controlled Trial Results to a Target

Population

Benjamin A. Goldstein, Matthew Phelan, Neha J. Pagidipati, Rury R. Holman, Michael J. Pencina

Elizabeth A Stuart

Benjamin A. Goldstein

Department of Biostatistics and Bioinformatics

2424 Erwin Road,

Durham, NC  27705

ben.goldstein@duke.edu

ACKNOWLEDGMENTS

We thank the NAVIGATOR steering committee and investigators for access to the NAVIGATOR data

Affiliations: Department of Biostatistics & Bioinformatics, Duke University, Durham, NC (BAG,

MJP); Center For Predictive Medicine, Duke Clinical Research Institute, Durham, NC (BAG, MP,

NHJ); Department of Medicine, Duke University, Durham, NC (NHJ); Diabetes Trials Unit, Oxford

Centre for Diabetes, Endocrinology, and Metabolism, University of Oxford, Oxford (RRH);

Department of Biostatistics, Johns Hopkins University, Baltimore, MD (EAS)

Funding: This work was supported by National Institute of Diabetes and Digestive and Kidney

Diseases (NIDDK) career development award K25 DK097279 (B.A.G.), US Department of Education

Institute of Education Sciences Grant R305D150003 (EAS). The project described was supported by the National Center for Advancing Translational Sciences (NCATS), National Institutes of Health (NIH), through Grant Award Number UL1TR001117 at Duke University. The content is solely the responsibility of the authors and does not necessarily represent the official views of the NIH. NAVIGATOR was funded by Novartis.

An Outcome Model Approach to Translating a Randomized Controlled Trial Results to a Target Population

Abstract

Participants enrolled into randomized controlled trials (RCTs) often do not reflect real-world populations. Previous research in how best to translate RCT results to target populations has focused on weighting RCT data to look like the target data. Simulation work, however, has suggested that an outcome model approach may be preferable. Here we describe such an approach using source data from the 2x2 factorial NAVIGATOR trial which evaluated the impact of valsartan and nateglinide on cardiovascular outcomes and new-onset diabetes in a "pre-diabetic" population. Our target data consisted of people with "pre-diabetes" serviced at our institution. We used Random Survival Forests to develop separate outcome models for each of the 4 treatments, estimating the 5-year risk difference for progression to diabetes and estimated the treatment effect in our local patient populations, as well as sub-populations, and the results compared to the traditional weighting approach. Our models suggested that the treatment effect for valsartan in our patient population was the same as in the trial, whereas for nateglinide treatment effect was stronger than observed in the original trial. Our effect estimates were more efficient than the weighting approach.

**Introduction**

Given good treatment compliance and minimal loss to follow-up, randomized controlled trials (RCTs) provide an internally valid estimate of a sample average treatment effect (SATE) for the evaluated intervention. If the RCT patient cohort is representative of the larger patient population, one can typically extrapolate the SATE to the larger patient population, to give a population average treatment effect (PATE). Unfortunately, for both intentional and unintentional reasons, RCTs rarely fully reflect general patient populations(1–3). This may be because it is advantageous to recruit patients that are not taking other medications, have fewer comorbidities or who are more likely to experience the primary outcome of interest. If the effectiveness of the intervention varies based on factors that differ between the RCT and general population, i.e. there are effect modifications, then the PATE will not equal the SATE and the inference derived from an RCT may not be valid in different clinical populations.

For these reasons, there has been increasing interest within the causal inference literature in developing methods to translate RCT results to target populations, estimating what is referred to as the target average treatment effect (TATE)  (4–11). The majority of this work has aimed to account for the selection process into the trial by developing weights using a model of the probability of a person being as part of the RCT versus the target population. These weights are then used to make the RCT population look like the target population – typically a larger national population. These approaches have good theoretical properties and have been developed to incorporate a double-robust framework(12).

The weighting approach, however, has two potential drawbacks. First, every time results need to be translated to a new patient population, a new set of weights needs to be estimated.  In most of the literature the target population is taken as the general US population, e.g. defined by NHANES, but in our work we often consider the target population to be patients served by the health system at a local institution. Amid changes to patient reimbursement, medical centers are becoming financially responsible for managing the health of their patient population (13).In order to manage a population cost effectively, health systems need to be able to reliably translate RCT results to real world populations. Since each health system has different patient characteristics, any application to any new health system, or population subset of interest, would require an estimation of a new set of weights. Second, the ability to estimate these weights accurately is partially driven by the sample size of the target population. When generalizing to a single, large population, this is not a concern. However, there may be situations where the local population is small, e.g. a single hospital or clinic. Accordingly, methods that are not dependent on the target sample size are desirable.

With these weighting approach limitations in mind, we have considered an outcome model approach to generalizing RCT results. The reasoning behind the outcome model approach, described in detail below, is that prediction models are built among those receiving or not receiving the intervention and then the target population is "passed" through each model to produce potential intervention outcomes for each individual in the target population. This methodology allows individual treatment effects (ITEs) to be estimated which can be averaged to calculate the TATE. This approach resolves the two challenges itemized above: the model only needs to be estimated once before applying to any target population, and that target population can be as small as desired. Kern et al.(11) used simulation to compare weighting, doubly robust, and outcome model approaches and found that the outcome model approach had the best performance.

In this paper we build off Kern ozan aet al. to further develop a machine learning approach to estimate the TATE. To do so we incorporate the causal Random Forests (RF) framework (14). Our intent is not to argue for supremacy of the outcomes model approach over the weighting approach but moreso to illustrate how to implement it. We used source data from the 2x2 factorial-design NAVIGATOR (Nateglinide And Valsartan in Impaired Glucose Tolerance Outcomes Research) international trial which evaluated the impact of valsartan and nateglinide on cardiovascular outcomes and new-onset diabetes in 9,306 "pre-diabetic" individuals. We then applied our results to "pre-diabetic" individuals found in our institution's electronic health record (EHR) system to estimate the TATE for this local patient population.

METHODS

We first describe the general analytic approach. Next we describe the data used for analysis, both source and target. Finally we outline our evaluation.

Analytic Approach

Figure 1 presents a schematic oozan af the analytic approach. Our approach builds off of work by Lu et al. (14) on causal RF. RFs (15) are an extension of Categorization and Regression Trees (CART) that combines multiple trees via a process called *bagging* (bootstrap aggregation) to create a more robust predictor. RF is a highly effective prediction model that has been used in a range of clinical studies, and is increasingly being used for causal inference (16). Lu et al. describe seven different approaches for estimating the ITE with RF. The approach we adopt here is referred to as *Counterfactual RF* (cRF). In cRF one identifies an outcome Y, treatments T ∈ {A,B} and covariates W. One then splits the data between those that received the treatment A, and those that received treatment B. Using the covariates W, one builds a model for the outcome separately for each treatment group. This generates two models, m:

$$\widehat{m_a} \equiv E\left(Y \vee W_s, T=A\right); \widehat{m_B} \equiv E\left(Y \vee W_s, T=B\right)$$

To estimate the ITE for a new observation, e.g. someone from a target sample, one passes the observation through each model, generating predicted values under each condition:

$$\widehat{Y_{i,A}} = m_A\left(W_{T,A}\right); \widehat{Y_{i,B}} = m_B\left(W_{T,B}\right)$$

We define the ITE as:

$$\hat{\tau}_i = \widehat{Y_{i,A}} - \widehat{Y_{i,B}}$$

We then average over all $\hat{\tau}_i$ to get the TATE. We note that this is a modification of the approach by Lu et al., who utilize the out-of-bag sample of RF to estimate the ITE within the developmental data. The rationale behind the approach is that by generating two separate models, any heterogeneity is implicitly modeled, allowing the outcomes under each condition to freely differ. Moreover, as we illustrate in our application, when there are multiple potential treatments, multiple contrasts is a straightforward calculation of the relevant comparison. An implicit assumption in this approach is that the potential outcomes for an individual are independent of one another, given the observed covariates. If there are unobserved effect modifiers any estimation of the TATE will produce biased results.

To generate standard errors for the TATE we need to determine the source of variability. The above procedure is comprised of two set of data: a RCT source sample and a local target patient population. We use the RCT data to estimate $\widehat{m()}$. We consider this sample random and consequently $\widehat{m()}$ random. Conversely, we consider the target population fixed, and the TATE, once $\widehat{m()}$ is estimated, fixed. Therefore, the primary source of variability comes from the estimation of $\widehat{m()}$. To estimate this variability, we follow the approach of Lu et al. and generate bootstrap samples of the data, refitting the RF models on each bootstrap and reestimating the individual ITEs and combined TATE. We use the estimated standard errors from the bootstrap distribution along with a normal approximation to generate a confidence interval. By generating standard errors in this way, the variability is not a function of the target sample size, only the source sample size.

Aligning the Input Data

An implicit component of this process is that the same W exist within both the source and target samples. Typically, RCTs have dozens of baseline covariates that are well defined and adjudicated. Conversely, EHR data typically have hundreds of covariate values that are not necessarily available for all patients. Moreover, similar measures are not necessarily equally defined. For example, within an

RCT a glucose test result may be measured via fasting glucose while an EHR may contain a mixture of fasting and random glucose tests. As such many of the same data elements may not exists in both data sources(17).  Therefore, care is necessary to ensure that there is alignment between the input variables. Particularly, it may be necessary to remove W that are not present in both data sources, as we describe below.

Data

*Source Data.* For our analysis we used data from the NAVIGATOR trial(18). NAVIGATOR was a 2x2 factorial-design trial comparing two medications, valsartan and nateglinide, in people with "pre-diabetes. These medications were compared against each other, against a placebo, and in combination. In total there were six comparisons. Published results found that Valsartan(19) was effective in reducing the incidence of diabetes while Nateglinide(20) was not.

In the trial, of the 9,306 participants, 2315, 2329, 2316, and 2346 were allocated to receive valsartan monotherapy, nateglinide monotherapy, valsartan-nateglinide combination therapy, and placebo, respectively. Inclusion criteria included impaired glucose tolerance (defined by glucose levels ≥140 mg/dL (7.8 mmol/L) but <200 mg/dL (<11.1 mmol/L) 2 hours after 75g glucose intake), plus one or more cardiovascular risk factors including: family history of premature coronary heart disease, current smoker, hypertension, reduced high-density lipoprotein, elevated low-density lipoprotein, left ventricular hypertrophy, microalbuminuria. Exclusion criteria included the use of an ACE inhibitor or ARB for the treatment of hypertension or the use of an antidiabetic medication in the last 5 years. Baseline information was available across 46 clinical and demographic factors. Median follow-up was 6.5 years.  We considered our primary end-point to be the risk difference for new-onset diabetes at 5 years.

*Target Data.* Our motivation for this work was to translate trial results to our institution's patient population. Duke University Health System (DUHS) consists of three hospitals and a network of outpatient clinics, along with linkable data from a federally qualified health clinic, serving an under-served population. As the only provider in Durham County, it is estimated that 85% of Durham County residents receive their primary care through DUHS(21).

To identify the local, target population we abstracted data from our institution's EHR system (See Figure 2 for Consort Diagram). To identify pre-diabetics we selected patients with a glycated hemoglobin A1C between 5.7% and 6.4%, with the condition that they never had a previous result greater or equal to 6.5%. We considered the first encounter with an HgB A1C in this range as the index date. We included patients seen between 2010 – 2016. To ensure that DUHS was a patient's medical home, we limited our analysis to individuals that lived in Durham County and had at least 2 encounters in the 2 years prior to the index date.

In total we were able to identify 35 (76%) of the NAVIGATOR baseline characteristics within our EHR. This included demographic variables, vital signs, comorbidities, labs and medications.  For variables with multiple measurements (e.g., systolic blood pressure) we used the median of all measurements taken in the year prior to the index date. Variables that we were not able to confidently identify within the EHR included: current smoking status (though smoking history was available), familial diabetes history, waist circumference, plasma glucose 2 hr after glucose load, and ratio of urinary albumin to creatinine. Since the occurrence of fasting-glucose was rare within our EHR, we used glycated hemoglobin A1C as an indicator of a pre-diabetic patient.

Analytic Approach

*Preliminary Analyses.*  We compared the patient characteristics between the NAVIGATOR and DUHS patients using standardized mean differences (SMD). We consider an SMD greater 0.1 to indicate a meaningful difference between the two populations(22).

*Primary Analyses.* We performed a series of analyses considering all six components of the 2x2 factorial design, but focusing on the valsartan and nataglinide comparisons versus placebo. We first estimated the SATE in the original NAVIGATOR RCT, estimating the risk difference at 5 years for each of the six comparisons. We performed this analysis as an intent-to-treat analysis fitting a Kaplan

-Meier estimator and then taking the effect estimate at 5-years. Next, we used the NAVIGATOR data to build a cRF for the valsartan, nataglinide, combination therapy and placebo cohorts. Specifically, we used Random Survival Forests(23) to predict risk of diabetes at 5-years for each arm. We fit 500 trees within each forest. Using these forests we considered the six potential contrasts. We first used the forests to estimate the treatment effect among the NAVIGATOR sample. We used the out-of-bag sample from each bootstrap iteration to estimate the probability of diabetes, and then, for each contrast of interest, calculated the ITE and TATE as described above. We considered this analysis a test of internal validity of the method, i.e. this analysis should replicate the estimate in the first analysis. Third, we applied the pre-diabetic individuals in the DUHS population to the cRF to estimate the target sample risk difference. We estimated the risk difference as above. For each analysis we performed 1000 bootstraps to estimate standard errors and calculate 95% confidence intervals.

As a secondary analysis, we considered the TATE across different sub-populations. Specifically, we considered differences based on: Sex, Race, and history of CVD.

*Sensitivity Analysis.* We performed two sensitivity analyses. First we limited the cohort to those individuals who would have met inclusion criteria for the original Navigator trial. Specifically, we required individuals to have at least one cardiovascular risk factor (history of smoking, hypertension, left ventricular hypertrophy, microalbumineria, reduced HDL [HDL < 40] or elevated LDL [LDL > 160]), excluded anyone currently on an ACE or ARB and limited the cohort to 2010 – 2011. Second, we compared our analysis to the more traditional weighting approach. While different proposals have been made, we followed the approach outlined in (5). We used the baseline characteristics to estimate selection weights. We note the authors mention using subject matter knowledge to decide which characteristics to include. Given our sample size we used all available baseline variables, using a RF estimated probability weights. We then performed a weighted regression to estimate the risk difference, and 1000 bootstraps to generate standard errors.

All analyses were performed in R 3.4.2. The RF model was estimated using the package randomSurvivalForests (24). This work was approved by our institution's IRB.

RESULTS

We identified 20,068 pre-diabetic patients in the DUHS EHR system. Table 1 shows the comparison of the source RCT and target EHR populations. In general, the RCT population was sicker with higher systolic blood pressure, more comorbidities, and more medication prescriptions. Of note the two groups had comparable HgbA1c, an important marker for pre-diabetes.

Assessment of Valsartan & Nateglinide

We first assessed the independent effect of Valsartan versus placebo (Table 2). Using the original NAVIGATOR data we estimated a significant effect for the risk difference at 5-years (-0.056, 95% CI: -0.085, -0.027), as in the originally published RCT. We were able to replicate this result when re-translating the cRF result back to the NAVIGATOR patient population (-0.051, 95% CI: -0.073, -0.028). Finally, when generalizing the effect to the Duke EHR defined population we get a similar, if not slightly stronger, effect estimate (-0.069, 95% CI: -0.119, -0.016).

We next assessed the independent effect of Nateglinide vs placebo. The original RCT reported a null effect, which we were able to confirm using the source data (0.009 95% CI: -0.021, 0.039), as well as through our internal validation of generalizing back to the RCT sample (-0.004 95%CI: -0.02, 0.013). However, when translating the effect to Duke EHR defined population we get a stronger, if not statistically significant effect (-0.030 95% CI: -0.077, 0.016).

We also assessed the additional comparisons of the two active drugs with each other as well as combination therapy versus mono-therapy (Table 2). Of note, the original RCT found valsartan to be superior to nateglinide, which was confirmed in our re-analysis of the risk difference (-0.065, 95% CI: -0.094, -0.036). However when translating the effect to the Duke patient population, this effect is attenuated and is no longer statistically significant (-0.039, 95% CI: -0.085, 0.010), likely due to the larger effect size of nateglinide versus placebo.

Subgroup Analyses

To illustrate the flexibility of the approach we examined the treatment effect in different sub-populations of our target populations (Figure 3). In general, we found similar treatment effects among the different sub-populations. One notable exception was the treatment effect for nateglinide among Caucasians (-0.016, 95% CI: -0.051, 0.021) versus African Americans (-0.041 95% CI: -0.095, 0.012). Since there is a greater proportion of African Americans in the Duke patient population, compared to the NAVIGATOR sample, it is possible that this difference is what accounts for the overall different observed treatment effect for nateglinide.

Sensitivity Analysis

As a sensitivity analysis we restricted the cohort to those patients that could have met inclusion criteria for the original NAVIGATOR study (n = 3,952). The results were very similar to the full sample results (Table 2). We next compared our findings to the more commonly applied sample weighting approach (Table 2). We found translated treatment effects of –0.054 (95% CI: -0.131, 0.021) and 0.008 (-0.080, 0.089), for the valsartan vs placebo and nateglinide vs placebo comparisons respectively. While these effects are more similar to the original trial estimates, we note the particularly large standard-errors. The average standard-error for the outcome model were 0.02 versus 0.04 for the weighting approach.

DISCUSSION

We have illustrated an approach to translate a treatment effect from an RCT to a target population. In contrast to much of the literature in this area, which focuses on weighting the source data, we focus on developing an outcome model. This allows us to treat this as two step process, decoupling the source and target populations results in two key advantages. First, we are able to easily translate the result to any population or sub-population of interest using the model fit in the trial sample. We do not need to reestimate weights for each new target. Second, the ability to translate the results is not dependent on the sample size of the target population.

In comparing the RCT sample to our local patient population, we noted that the RCT population was generally sicker, with higher blood pressure, more comorbidities and more medications. This confirms previous literature that has shown that RCT samples are generally sicker than the general population(3). Interestingly, other work has suggested that EHR based populations are sicker than general clinical populations (25) suggesting that these differences may be an underestimate, and in fact differences may be even more extreme when compared to a general patient population.

Upon translating the RCT results to our local population we were able to find a similar, significant, TATE as the RCT based SATE for the valsartan vs placebo comparison. However, while the SATE for the nateglinide vs placebo comparison showed no treatment benefit, the TATE did suggest some treatment effect – though not statistically significant. This result followed through when we considered some of the comparisons based on combination therapy. For example, the SATE for valsartann + nateglinide versus just nataglinide showed a significant treatment effect, while the TATE did not.

An important feature of the outcome model approach is that the estimation of the outcome model is decoupled from the translation step. This means that an RCT can generate an outcome model without knowledge of the target population. All that is required is to be able to map the covariates in the target population back to the trial data – admittedly not always an easy task. This feature is what makes translation to additional sub-populations so efficient. When we applied our analysis to different sub-populations we found similar treatment effects within the exception of a heterogeneous racial effect among those taking nateglinide. While it is typical to use trial data to assess treatment heterogeneity, we note that the original trial did not report a significant effect for racial heterogeneity (p = 0.82). This suggests that this may effect differences is not due solely to racial differences, but higher-order interaction effects. Overall, it is likely that these differences is what accounted for the difference in the SATE and TATE estimates.

Embedded in our analysis is the ability to check the internal validity of the outcome model. By leveraging the out-of-bag samples from the RF bootstrap iterations, we re-applied the outcome model to the RCT population. We were able to find similar effect estimates as the original trial, suggesting that our approach is relatively stable. We should suggest that this should be a standard first step when applying such an analysis.

In addition to the computational efficiency of using the out-of-bag sample to validate the approach, there are other advantages to using RF to derive the outcome model. Since RF is based on trees, it is more robust to outlying values, and consequently extrapolation, from the target sample. Unlike linear models, trees do not extrapolate predictions when the observed covariates extend beyond the support of the source data. This is an important consideration because, as we noted, there were meaningful differences between the RCT and DUHS samples. In general the question of how to estimate the TATE when the source and target samples differ is worthy of more future research.

Finally, we note that when we compared our outcome model approach to the more typical weighting approach we found our effect estimates to have meaningfully smaller standard errors. This empirical finding confirms the simulation results of Kern et al.(11) One intuition for why this is the case is that the estimation of sample weights relies on both the target and source sample while the proposed approach only relies on the source data. Moreover, RF is a relatively stable predictor that has low predictive variance.(26)

There are some notable limitations in our analysis and analytic approach. First, these are the results of one analysis. This approach should be tried with different source and target samples to note any additional potential complications as well as additional simulation work to better understand when and how the methods work. It would be interesting to compare the TATE different in different clinical populations to assess how much potential clinical heterogeneity exists. Moreover in our analysis we had a relatively large RCT from which to work. It is worth investigating what is the minimum size under which such an approach will still provide valid and efficient estimates. More generally, additional work in this field should consider the effects of translating trial results outside the support of the trial data. Finally, a key challenge in this area is what to do when variables are not directly comparable between the source and target data. This is an especially important concern with EHR data where clinical factors are measured in different ways. Hong et al.(6) has done some work in this area showing how multiple imputation can be used for missing individual values, but more work is needed to understand how best to handle, observed, but unmappable, covariates.

In this analysis, we illustrate a means to translate a RCT result to a target population. By using an outcome model approach, we are able to efficiently translate the results to target populations of different sizes.

| Characteristic | Local Pre-Diabetic Population (N=20,068) | NAVIGATOR (N=9,306) | Standardized Mean Difference |
|---|---|---|---|
| **Table 1 - Differences in Baseline Characteristics of Target and Source Populations** | | | |
| Age (yrs) (Median, Percentiles) | 52.6 (41.6- 63.2) | 63.0 (58.0- 69.0) | -0.907 |
| Female sex | 12601 (62.8%) | 4711 (50.6%) | 0.247 |
| Race | | | |
|     Asian | 656 (3.3%) | 613 (6.6%) | -0.154 |
|     Black | 9927 (49.5%) | 236 (2.5%) | 1.266 |
|     Other | 2245 (11.2%) | 723 (7.8%) | 0.117 |
|     White | 7240 (36.1%) | 7734 (83.1%) | -1.092 |
| Weight (kgs) (Median, Percentiles) | 87.8 (74.2- 104) | 82.0 (71.5- 93.5) | 0.148 |
| BMI (Median, Percentiles) | 31.4 (27.1- 37.1) | 29.7 (26.8- 33.3) | 0.046 |
| SBP (mmHg) (Median, Percentiles) | 128 (119- 138) | 140 (128- 150) | -0.588 |
| DBP (mmHg) (Median, Percentiles) | 78.3 (72.0- 84.4) | 84.0 (78.0- 90.0) | -0.57 |
| | | | |
| **Cardiovascular Risk Factors** | | | |
| Any | 16502 (82.2%) | 8921 (95.9%) | -0.447 |
| Smoking History | 7497 (37.4%) | 1025 (11.0%) | 0.647 |
| Hypertension | 12016 (59.9%) | 7216 (77.5%) | -0.388 |
| Left Ventricular Hypertrophy | 968 (4.8%) | 268 (2.9%) | 0.101 |
| Microalbuminuria | 244 (1.2%) | 114 (1.2%) | -0.001 |
| | | | |
| **History of cardiovascular disease** | | | |
| Any | 2651 (13.2%) | 2745 (29.5%) | -0.406 |
| Myocardial Infarction | 504 (2.5%) | 1103 (11.9%) | -0.368 |
| Angina | 441 (2.2%) | 1561 (16.8%) | -0.514 |
| Percutaneous Coronary Intervention | 777 (3.9%) | 622 (6.7%) | -0.126 |
| Coronary-Artery Bypass Grafting | 20 (0.1%) | 521 (5.6%) | -0.335 |
| Intermittent Claudification | 526 (2.6%) | 98 (1.1%) | 0.117 |
| Lower-limb angioplasty | 16 (0.1%) | 110 (1.2%) | -0.14 |
| Nontraumatic leg or foot amputation | 14 (0.1%) | 7 (0.1%) | -0.002 |
| Stroke | 1506 (7.5%) | 374 (4.0%) | 0.15 |
| | | | |
| **Labs at time of index date** | | | |
| Fasting Glucose mmol/L (Median, Percentiles) | 5.4 (5.0- 5.9) | 6.1 (5.7- 6.4) | -0.633 |

| Characteristic | Local Pre-Diabetic Population (N=20,068) | NAVIGATOR (N=9,306) | Standardized Mean Difference |
|---|---|---|---|
| **Table 1 - Differences in Baseline Characteristics of Target and Source Populations** | | | |
| A1C (Median, Percentiles) | 5.9 (5.8- 6.1) | 5.8 (5.5- 6.1) | 0.393 |
| Total Cholesterol (Median, Percentiles) | 187 (161- 214) | 207 (181- 236) | -0.503 |
| HDL-C (Median, Percentiles) | 46 (38- 55) | 48 (40- 57) | -0.132 |
| LDL-C (Median, Percentiles) | 113 (91- 137) | 124 (100- 150) | -0.312 |
| Creatinine mg/dL(Median, Percentiles) | 0.9 (0.8- 1.0) | 0.8 (0.7- 1.0) | 0.184 |
| Estimated GFR (Median, Percentiles) | 90 (74- 106) | 80 (69- 91) | 0.194 |
| eGFR<60 | 1842 (10.8%) | 1025 (11.1%) | -0.061 |
| | | | |
| **Medications at time of index date** | | | |
| ACE inhibitor | 3346 (16.7%) | 676 (7.3%) | 0.293 |
| Angiotensin-receptor blocker | 1239 (6.2%) | 30 (0.3%) | 0.335 |
| Alpha-blocker | 633 (3.2%) | 577 (6.2%) | -0.145 |
| Aspirin or other antiplatelet drug | 5500 (27.4%) | 3425 (36.8%) | -0.202 |
| Beta-blocker | 3376 (16.8%) | 3666 (39.4%) | -0.518 |
| Calcium-channel blocker | 3354 (16.7%) | 3012 (32.4%) | -0.37 |
| Diuretic | 4895 (24.4%) | 2960 (31.8%) | -0.166 |
| Lipid-modulating drug | 6496 (32.4%) | 3577 (38.4%) | -0.127 |
| Any Medication | 11450 (57.1%) | 7794 (83.8%) | -0.612 |

**Table 2:** Risk difference at 5-years across different treatment comparisons and populations

| | Valsartan VS Placebo | Nateglinide VS Placebo | Valsartan VS Nateglinide | Valsartan + Nateglinide VS Placebo | Valsartan + Nateglinide VS Valsartan | Valsartan + Nateglinide VS Nateglinide |
|---|---|---|---|---|---|---|
| **Sample Average Treatment Effect in NAVIGATOR Trial** | -0.056 (-0.085 , -0.027) | 0.009 (-0.021 , 0.039) | -0.065 (-0.094 , -0.036) | -0.036 (-0.065 , -0.007) | 0.02 (-0.009 , 0.048) | -0.045 (-0.075 , -0.016) |
| **Re-Translation to NAVIGATOR Sample*** | -0.051 (0.073 , -0.028) | -0.004 (-0.02 , 0.013) | -0.047 (-0.064 , -0.032) | -0.047 (-0.063 , -0.031) | 0.004 (-0.012 , 0.021) | -0.043 (-0.067 , -0.021) |
| **Translation to local pre-diabetic population*** | -0.069 (-0.119 , -0.016) | -0.030 (-0.077 , 0.016) | -0.039 ( -0.085 , 0.01) | -0.043 (-0.086 , 0.004) | 0.025 (-0.025 , 0.072) | -0.013 (-0.056 , 0.03) |
| **Translation to local pre-diabetic population eligible for NAVIGATOR*** | -0.069 (-0.119 , -0.015) | -0.031 (-0.082 , 0.017) | -0.038 (-0.084 , 0.012) | -0.036 (-0.082 , 0.014) | 0.034 (-0.019 , 0.083) | -0.004 (-0.049 , 0.042) |
| **Translation to local pre-diabetic population using weighting to local population** | -0.063 (-0.147 , 0.017) | 0.003 (-0.091 , 0.094) | -0.06 (-0.152 , 0.022) | -0.03 (-0.121 , 0.067) | 0.031 (-0.056 , 0.111) | -0.03 (-0.116 , 0.06) |

**\* Each of these methods used the counter-factual random forests aproach for estimation**

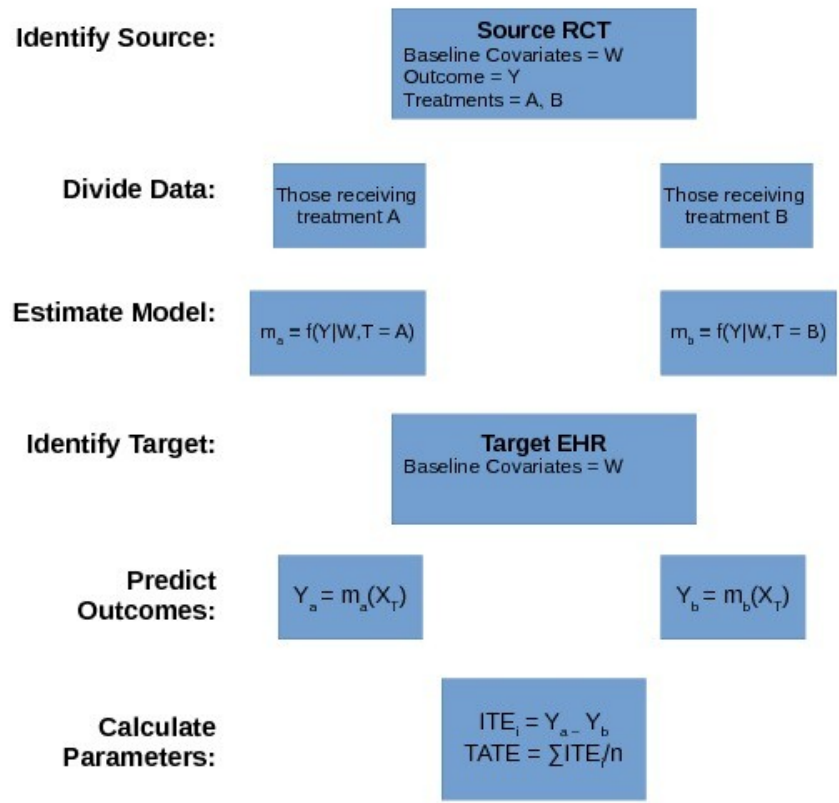

**Figure 1:** Flow diagram for implementation of outcome approach. We start with RCT Source data. Divide the data between those receiving the treatments of interest. Estimate separate outcome models. Identify the target data. Predict individual outcomes under each model. Calculate the individual treatment effect (ITE) and average to obtain the target averaged treatment effect (TATE).

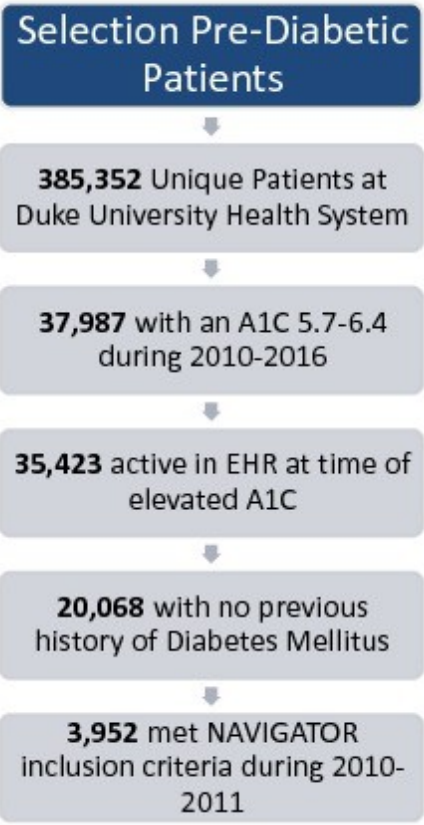

**Figure 2:** Consort diagram for identification of EHR based target cohort.

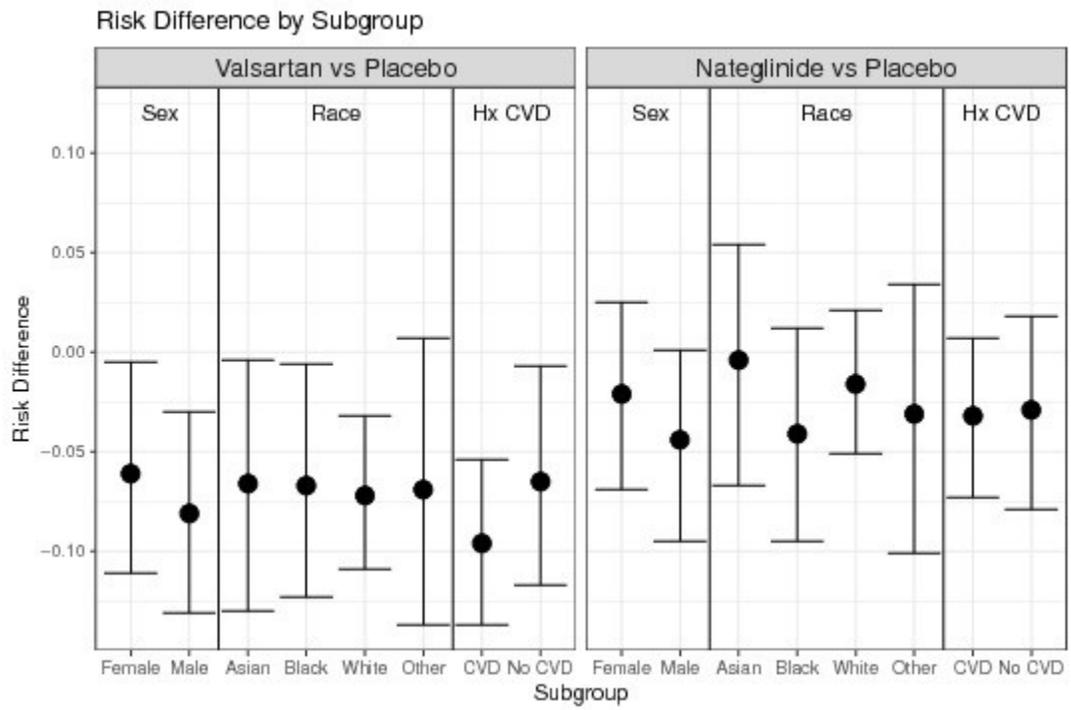

**Figure 3:** Effect estimates for different patient subgroups.